\documentclass[prb, amsmath, amssymb, reprint, showpacs, showkeys, footinbib]{revtex4-1}

\usepackage{graphicx}
\usepackage{dcolumn}
\usepackage{bm}
\usepackage{epstopdf} 

\usepackage{cellspace}
\usepackage[normalem]{ulem}
\usepackage{xcolor}

\begin{document}

\title{Local magnetic anisotropy by polarized neutron powder diffraction: application of magnetically induced preferred crystallite orientation}


\author{I.A.\,Kibalin$^{1, 2}$ and A.\,Gukasov$^{1}$}

\affiliation{$^1$Laboratoire L\'eon Brillouin, CEA-CNRS, CE-Saclay, 91191 Gif-sur-Yvette, France}
\affiliation{$^2$PNPI NRC "Kurchatov Institute" Orlova rosha, Gatchina, 188300 Leningrad region, Russia}
\email{arsen.goukassov@cea.fr}

\date{\today}

\begin{abstract}
Polarized neutron diffraction allows to determine the local susceptibility tensor on the magnetic site both in single crystals and powders.  It is widely  used in the studies of single crystals, but it is still hardly applicable to a number of highly interesting powder materials, like molecular magnets or nanoscale systems because of the low  luminosity of existing instruments and the absence of an appropriate data analysis software. We show that these difficulties can be overcome by using a large area detector in combination with the two-dimensional Rietveld method and powder samples with magnetically induced preferred crystallite orientation. This is demonstrated by revisiting two test powder compounds, namely,  low anisotropy (soft) ferrimagnetic compound Fe\textsubscript{3}O\textsubscript{4} and spin-ice compound Ho\textsubscript{2}Ti\textsubscript{2}O\textsubscript{7} with high local anisotropy. The values of magnetic moments in Fe\textsubscript{3}O\textsubscript{4} and the susceptibility tensors of Ho\textsubscript{2}Ti\textsubscript{2}O\textsubscript{7} at various temperatures and fields were found in perfect agreement with these found earlier in single crystal experiments. 
The magnetically induced preferred crystallite orientation was used to study the local susceptibility of a single-molecule magnet Co([(CH\textsubscript{3})\textsubscript{2}N]\textsubscript{2}CS)\textsubscript{2}Cl\textsubscript{2}. Hence, the studies of local magnetic anisotropy in powder systems might now become accessible.

\end{abstract}


\maketitle

\section{Introduction}

Polarized neutron diffraction (PND), also called ``flipping ratio method'' is a powerful tool to investigate intra- or intermolecular magnetic interactions. It gives a direct access to the magnetization distribution in the unit cell\cite{Gillon_2012}, permits separating the spin and orbital contributions\cite{Schweizer_2008} and allows  determining the local susceptibility tensor on the magnetic sites\cite{Gukasov_2002_jpcm}. The magnetization distribution has contributed to the understanding of magnetic interactions by revealing the spin delocalization, the spin density distribution, and the wave functions of unpaired electrons\cite{Kibalin_2017}. In turn, the local susceptibility approach has been successfully used in recent studies of field induced magnetic order in R\textsubscript{2}Ti\textsubscript{2}O\textsubscript{7} pyrochlore compounds with either uniaxial or planar anisotropy\cite{Cao_2008, Cao_2009}. PND is becoming a reference in mapping the magnetic anisotropy at the atomic scale in molecular magnets\cite{Ridier_2015, Klahn_2018}. Unfortunately PND currently applies only to single crystals, which makes it inadequate for a number of highly interesting topics due to the difficulties encountered in growing sufficiently large samples. 

Motivated by challenging scientific subjects, several attempts have been performed to investigate magnetized powder samples with polarized neutrons\cite{LelievreBerna_2010, Wills_2005, Hiraka2014, Rivin_2013}. This allowed to reveal magnetic moments of iron at different crystallographic sites in prussian blue\cite{Wills_2005} and in $\alpha$-Fe\textsubscript{16}N\textsubscript{2} nanoparticles\cite{Hiraka2014}, as well as, to amend the magnetic structure of highly anisotropic TbCo\textsubscript{2}Ni\textsubscript{3}\cite{Rivin_2013}. The validity of the method was illustrated by measurements of magnetic anisotropy in a polycrystalline sample of Tb\textsubscript{2}Sn\textsubscript{2}O\textsubscript{7}\cite{Gukasov_2010}. As a proof of concept the local susceptibility parameters of Tb were found by a two steps procedure. First, the integrated intensities of the spin-up and spin-down components $I_{+}(hkl)$ and $I_{-}(hkl)$ were obtained by profile matching from the corresponding powder patterns. Then, the program CHILSQ of the Cambridge Crystallography Subroutine Library\cite{Brown_1993} was used to fit the integrated intensities. It is clear that such a procedure of data treatment can be applied only to highly symmetric crystal structures with small unit cell. For more complex structures a Rietveld method needs to be developed.

Rietveld analysis has become mandatory in powder diffraction for nuclear and magnetic structure refinement\cite{Rodriguez-Carvajal1993, Larson_1994, Petricek_2014}. It refines various metrics, including lattice parameters, structure and magnetic parameters, a preferred orientation to derive a calculated diffraction pattern. Once the calculated pattern becomes nearly identical to an experimental one, various properties pertaining to that sample can be obtained. However, the Rietveld method for polarized neutron powder diffraction (PNPD) has not been implemented yet. In the above-mentioned  polarized powder experiments  special softwares (model dependent) were developed for the data treatment.

We note that at first PNPD measurements were performed on conventional powder diffractometers equipped by one-dimensional (1D) detectors while modern unpolarized neutron powder diffractometers (Super-D2B, D20, SPODI) at reactor sources are equipped with two-dimensional (2D) detectors. Area detectors increase the efficiency of the instrument by an order of magnitude, but the common approach at these instruments consists in reducing the accumulated 2D data from area detector into 1D diffraction pattern by ``unbending'' measured Debye cones. The resulting pattern is then treated using standard 1D Rietveld refinement\cite{Petricek_2014, Rodriguez-Carvajal1993, Larson_1994}. Most recent powder diffractometers at advanced neutron spallation sources (WISH, POWGEN)  use very large area detectors and operate in TOF mode. This generates rather complex three-dimensional (3D) angular- and wavelength-dispersive data which are eventually transformed into one-dimensional diffraction pattern\,I(2$\theta$) (or I($\lambda$))\cite{mantid} to allow standard Rietveld refinement. It has been noted that two-dimensional extension of Rietveld method for neutron TOF powder diffraction taking into account the variation of diffraction angle\,2$\theta$ and wavelength\,$\lambda$ decreases the number of data-reduction steps and avoids the loss of high-resolution information \cite{Jacobs_2015}, but full scale multidimensional Rietveld software for neutron TOF powder diffraction still need to be developed.

Since  area detectors increase considerably the efficiency of the instruments, we performed our PNPD measurements on diffractometers  equipped with large 2D position sensitive detectors. When using  area detectors, the polarized neutron scattering is a function of\,2$\theta$ and $\varphi$ but also of the angle between the magnetic field and the scattering vector. Moreover, neutrons are sensitive only to the magnetic moment perpendicular to the scattering vector. Therefore, the variation of intensity along the Debye cones can be used for the separation of nuclear and magnetic scattering contributions. For these reasons the transformation of angular-dispersive polarized neutron data from area detectors into one-dimensional 2$\theta$ pattern is not applicable. We note as well that the equation for powder averaging derived in the paper\cite{Gukasov_2010} is valid only for the vertical field and scattering in the horizontal plane. Here we give an expression for powder averaging valid for general scattering geometry, which allows an implementation of the full scale 2D Rietveld method in PNPD.

Another possibility of increasing the efficiency of PNPD consists in using  magnetically induced  preferred crystallite orientation.
This technique can be applied to biaxial crystals in which the magnetic susceptibility tensor has different principal values (i. e. orthorhombic, monoclinic and triclinic systems)\cite{Kimura_2014, Stekiel2015}.
Under a strong magnetic field the crystallites overcome the steric hindrence of powder packing and align their easy magnetization axis parallel to the applied magnetic field, leading to crystallite preferred orientation. As a consequence, different reflections with similar Bragg angles $2\theta$ appear at different  angles along the Debye cones. No overlapping of these reflections occurs, which allows to use diffractometers with low resolution (hence, high luminosity)   for powder diffraction. 

Here we show that the combination of a large area detector with 2D Rietveld analysis and  magnetically induced  preferred crystallite orientation enables PNPD  in systems not available as single crystals.
We illustrate this by the results of  two test cases of magnetic materials; low anisotropy (soft) ferrimagnetic compound Fe\textsubscript{3}O\textsubscript{4} and spin-ice compound Ho\textsubscript{2}Ti\textsubscript{2}O\textsubscript{7} with high local anisotropy. We show that in both cases the combination of area detector with 2D Rietveld method shortens the acquisition time by an order of magnitude, without loosing the precision of parameters evaluation. Finally, we present the results of the local susceptibility studies on the single-molecule magnet Co([(CH\textsubscript{3})\textsubscript{2}N]\textsubscript{2}CS)\textsubscript{2}Cl\textsubscript{2}  with magnetically induced preferred orientation of crystallites, which shows that the PNPD  now opens large opportunities in the local anisotropy quantification of complex structures.

\section{Polarized neutron powder diffraction}

\begin{figure}
\includegraphics[width=7cm]{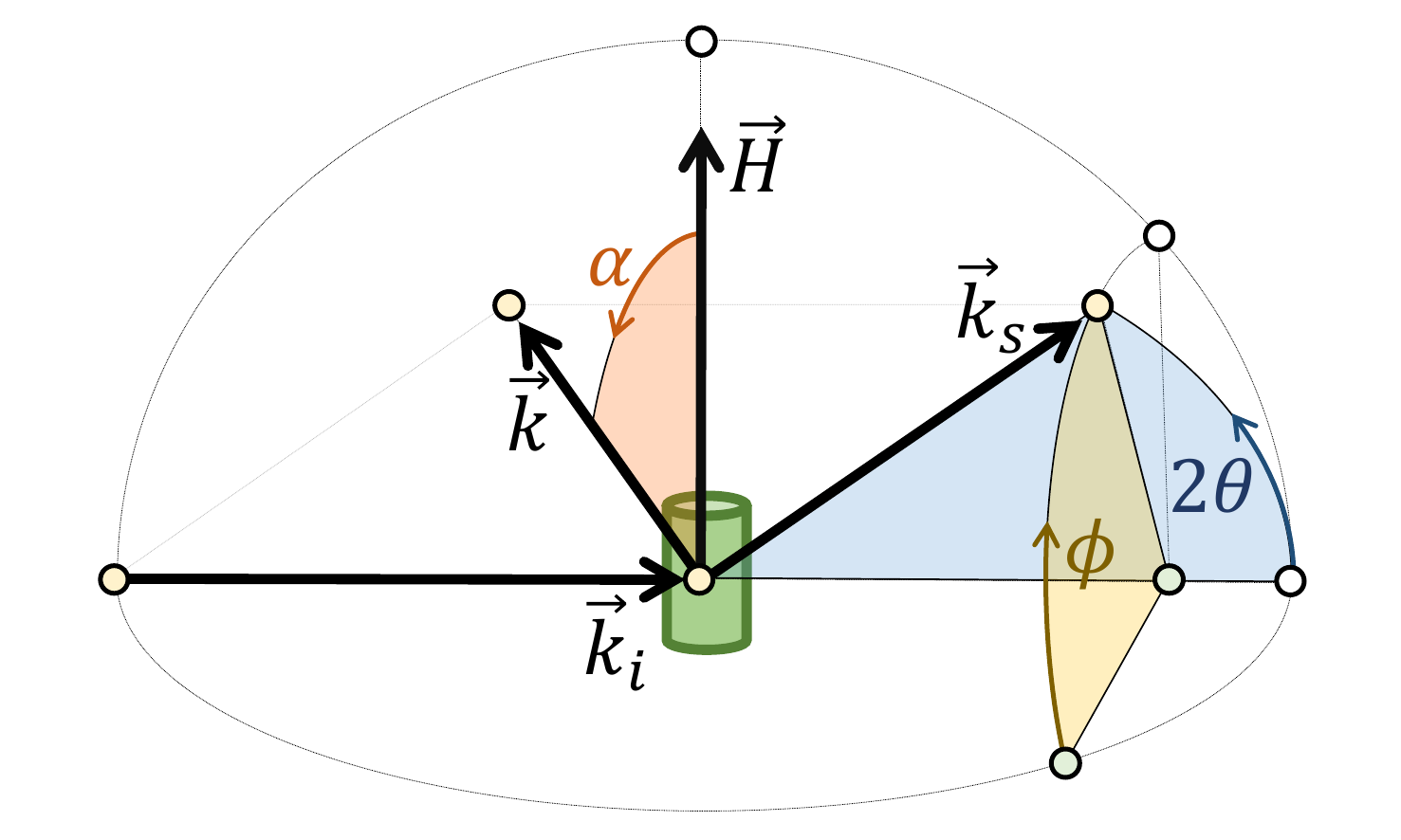}

\caption{(Color online) The principal scheme of the scattering at the pollycrystalline sample. A notation is explained in the text. \label{fig:For-different-crystallites}}
\end{figure}

It is well established that the flipping sum and difference of the integrated intensities ($I_{+}$ and $I_{-}$) of pollycrystalline samples are proportional to\cite{Gillon2007}

\begin{equation}
I_{+}+I_{-}\sim\left|N\right|^{2}+\left\langle \left|\vec{M}_{\perp}\right|^{2}\right\rangle ,\label{eq:flip_iint_sum}
\end{equation}

\begin{equation}
I_{+}-I_{-}\sim N^{*}\left\langle\left( \vec{M}_{\perp}\cdot\vec{P}\right)\right\rangle +N\left\langle \left(\vec{M}_{\perp}^{*}\cdot\vec{P}\right)\right\rangle,\label{eq:flip_iint_diff}
\end{equation}

\noindent
where $N$ is the nuclear structure factor, $\vec{M}_{\perp}$ is the projection of the magnetic structure factor $\vec{M}(\vec{k})$ perpendicular to the scattering vector $\vec{k}$. $\vec{M}$ is induced by the magnetic field $\vec{H}$ applied in the vertical direction (figure\,\ref{fig:For-different-crystallites}) and $\vec{P}$ is the neutron polarization vector parallel to $\vec{H}$. Angle brackets show the powder averaging over scattering crystallites.

In soft magnetic materials the atomic magnetic moments $\vec{M}_{a}$ are directed along the applied field $\vec{H}$. Thus, the powder averaging of $\left|\vec{M}_{\perp}\right|^{2}$, $\left( \vec{M}_{\perp}\cdot\vec{P}\right) $ can be written as:

\begin{equation}
\left\langle \left|\vec{M}_{\perp}\right|^{2}\right\rangle =\left|\sum_{a}M_{a}\sin\alpha f_{a}(\vec{k})\exp\left[2\pi i\vec{k}\cdot\vec{r}_{a}\right]\right|^{2}\label{eq:m_perp_sq_aver_smm}
\end{equation}

\noindent
and

\begin{equation}
\left\langle \left(\vec{M}_{\perp}\cdot\vec{P}\right)\right\rangle =\sum_{a} M_{a}P\sin^{2}\alpha f_{a}(\vec{k})\exp\left[2\pi i\vec{k}\cdot\vec{r}_{a}\right],\label{eq:m_perp_p_aver_smm}
\end{equation}

\noindent
where the sum over $a$ includes all atoms in the unit cell with radius vector $\vec{r}_{a}$, $f_{a}(\vec{k})$ is the magnetic form factor in spherical approximation\cite{IntTablesHB2004}.

For paramagnets and diamagnets the structure factor $\vec{M}$ can be written as\cite{Gukasov_2010}:

\begin{equation}
\vec{M}(\vec{k})=\sum_{a}\frac{1}{N_{a}}f_{a}(\vec{k})\sum_{p}R_{p}\chi_{a}R_{p}^{-1}\vec{H}e^{2\pi i\vec{k}(R_{p}\vec{r}_{a}+\vec{t}_{p})},\label{eq:SFT-1}
\end{equation}

\noindent
where the sum over $a$ includes all independent atoms, the sum over $p$ includes those generated from atom $a$ by the $N_{g}$ symmetry operators $\left\lbrace R_{p}:\vec{t}_{p}\right\rbrace$ of the space group $\mathcal{G}$. $N_{a}$ is the number of operators $q$ in $\mathcal{G}$ for which $R_{q}\vec{r}_{a}+\vec{t}_{q}=\vec{r}_{a}$; $N_{g}/N_{a}$ is the multiplicity  of the site $a$ which has point symmetry $Q_{a}$ generated by the rotational parts of the operators $q$. This implies that the local susceptibility tensor $\chi_{a}$ is invariant to the rotations in $Q_{a}$ so that $R_{q}\chi_{a}R_{q}^{-1}=\chi_{a}$ for all $R_{q}$ in $Q_{a}$. The number of independent components of the tensor $\chi_{a}$ varies from two for uniaxial site symmetries to six for triclinic ones.

The structure factor tensor $\chi(\vec{k})$, which is independent of the magnitude and direction of the applied field can be expressed as follows:

\begin{equation}
\chi(\vec{k})=\sum_{a}\frac{1}{N_{a}}f_{a}(\vec{k})\sum_{p}R_{p}\chi_{a}R_{p}^{-1}e^{2\pi i\vec{k}(R_{p}\vec{r}_{a}+\vec{t}_{p})}.\label{eq:SFT}
\end{equation}

Expression for the powder averaging of $\left|\vec{M}_{\perp}\right|^{2}$ and $\left(\vec{M}_{\perp}\cdot\vec{P}\right)$ terms in the case of magnetic field applied vertically and the detector in the horizontal plane has been given in the Ref.\cite{Gukasov_2010}. It can be shown (see Supplemental Material) that this expression can be generalized for any scattering geometry. Namely, the structure factor tensor is to be transformed into a Cartesian coordinate system with the $z$ axis parallel to the scattering vector. If the transformation is expressed through the matrix\,$T$ (see Supplemental Material) the components of the tensor become $\Sigma=T\cdot\chi\cdot T^{-1}$ and the averaged terms above can be written as follows

\begin{equation}
\begin{array}{c}
\left\langle \left|\vec{M}_{\perp}\right|^{2}\right\rangle =\frac{1}{2}H^{2}\left[\left(\Sigma_{11}^{2}+2\Sigma_{12}^{2}+\Sigma_{22}^{2}\right)\sin^{2}\alpha+\right.\\
\left.+2\left(\Sigma_{13}^{2}+\Sigma_{23}^{2}\right)\cos^{2}\alpha\right]
\end{array}\label{eq:m_perp_sq_aver}
\end{equation}

\begin{equation}
\left\langle \left( \vec{M}_{\perp}\cdot\vec{P}\right)\right\rangle =PH\left(\frac{\Sigma_{11}+\Sigma_{22}}{2}\right)\sin^{2}\alpha.\label{eq:m_perp_p_aver}
\end{equation}

\noindent
Here $\cos^{2}\alpha=\cos^{2}\theta\sin^{2}\phi$. These equations describe the scattering along the Debye cones in the 2D Rietveld refinement. We note that for the special case of scattering in the equatorial plane ($\phi=0$) the expressions (\ref{eq:m_perp_sq_aver}, \ref{eq:m_perp_p_aver}) are in exact accordance with these given before\cite{Gukasov_2010}.

\subsection{2D-diffraction profile}

In the two-dimensional case the calculated intensity $y_{\pm}(2\theta,\phi)$ for a single-phase diffraction pattern can be expressed for every data point by

\begin{equation}
y_{\pm}(2\theta,\phi)=S\sum_{h}m_{h}L_{f}P_{h}I_{\pm}(\alpha)\psi_{h}(2\theta-2\theta_{h},\phi)+b(2\theta,\phi),
\label{eq:profile}
\end{equation}

\noindent
where $S$ is a scale factor, $m_{h}$ is the multiplicity of reflection, $L(\theta,\phi)$ is the Lorentz factor, $P_{h}$ is the density of (hkl) poles at the scattering vector (preferred orientation), $\psi_{h}(2\theta-2\theta_{h},\phi)$ is the peak profile function normalized to unit area, and $b(2\theta,\phi)$ is the background. The summation is done over all $h$ reflections for each data point. For a cylindrical detector\cite{Norby_1997} the Lorentz factor is $\sqrt{1-\sin^{2}2\theta\sin^{2}\phi}/\sin^{2}\theta\cos\theta$. In the case of one-dimensional Rietveld refinement the profile function is usually described by the pseudo-Voight function. For the two-dimensional description of the diffraction pattern, an appropriate profile function still needs to be found. Here we used the standard one-dimensional expression for the profile function $\psi_{h}(2\theta-2\theta_{h})$ neglecting the dependence of the peak profile from the polar angle $\phi$ . For these reason the part of the diffraction pattern with strong dependence of the peak profile from the polar angle $\phi$ was excluded from the refinement procedure.

It has been suggested\cite{Hiraka2014, LelievreBerna_2010} that in the PNPD better quality information can be derived by using the flipping difference data, as contamination from the cryomagnet and sample is largely eliminated in the difference. However, we note that a simultaneous refinement of the sum and the difference patterns is mandatory for the scaling of magnetic moment values. Moreover, as has been noted in Ref.\cite{Gukasov_2010}, in the cases of strong magnetic scatters with high anisotropy the sum patterns might contain a number of purely magnetic reflections which do not depend on neutron polarization.

\subsection{Experiment and data treatment}

Neutron diffraction studies were performed at the Orph\'ee 14MW reactor of the Laboratory L\'eon Brillouin, CEA Saclay. The diffraction patterns were collected on the diffractometer 5C1, equipped by position sensitive detector with cylindrical geometry covering $80^{\circ}$ and $25^{\circ}$ in horizontal and vertical directions, using neutrons of wavelength $\lambda=0.84$\AA\, obtained with a Heusler alloy monochromator. The incident beam polarization $P$ is $0.91$.

PNPD data were collected on powder sample of Fe\textsubscript{3}O\textsubscript{4} in an external field of 0\,T and 6\,T below the Verwey transition (at 10\,K) and above it (at 150\,K). The experiments with sintered powder sample of Ho\textsubscript{2}Ti\textsubscript{2}O\textsubscript{7} were performed in the temperature range from 5\,K up to 50\,K in the magnetic field of 1\,T.

Measurements of the Co(II) complex with single-molecule magnet behavior have been carried out on the thermal polarized neutron lifting counter diffractometer 6T2 (LLB-Orph\'ee, Saclay). Neutrons of wavelength 1.4\,\AA\, were monochromated by a vertically focusing graphite crystal and polarized by a supermirror bender. The polarization factor of the beam was 0.95. The position sensitive detector has a flat geometry.  Data treatment was performed using the newly developed 2D Rietveld software RhoChi\footnote{The RhoChi source code is written in python\,3 using cryspy library. It is freely distributed through the GitHub service: https://github.com/ikibalin/cryspy, where a short guide of its application is given together with examples.}.

\section{Soft ferrimagnetic Fe\textsubscript{3}O\textsubscript{4}}

Magnetite Fe\textsubscript{3}O\textsubscript{4}, as an original magnetic material with modern applications ranging from spintronics to MRI contrast agents was chosen as an example of soft (low anisotropy) ferrimagnetic for the software benchmarking. At ambient temperatures it orders in inverted cubic spinel ferrite with the tetrahedral ($A$) site occupied by Fe\textsuperscript{3+} ions and with Fe\textsuperscript{2+} and Fe\textsuperscript{3+} ions coexisting at the same octahedral ($B$) site\cite{Hamilton_1958}. Magnetite undergoes a first-order transition below 120\,K where the resistivity increases by two orders of magnitude and the structural distortions from cubic symmetry occur\cite{Okamura_1932, Ellefson_1934}. It is suggested that this transition is driven by a charge ordering of Fe\textsuperscript{2+} and Fe\textsuperscript{3+} ions\cite{VERWEY_1939}. Polarized neutron diffraction measurements performed on the single crystal of magnetite earlier has shown the antiparallel orientation of the moments at the tetrahedral and octahedral sites but surprisingly no difference between the magnetic moments at the sites was found\cite{Rakhecha_1978}.

\begin{figure}
\includegraphics[width=8cm,height=4.36cm]{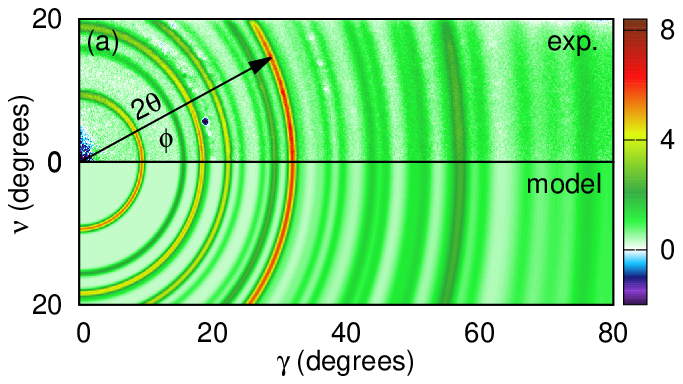}

\includegraphics[width=8cm,height=4.36cm]{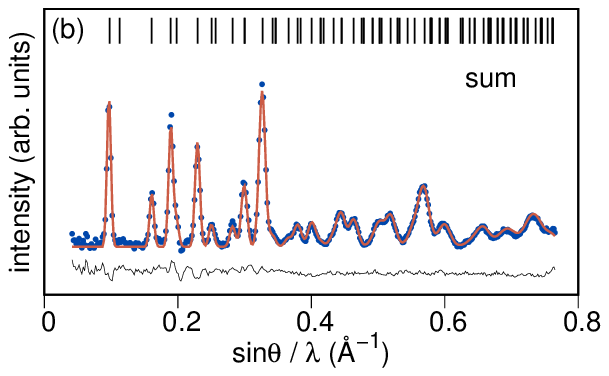}

\caption{(Color online) Flipping sum diffraction patterns collected on Fe\textsubscript{3}O\textsubscript{4} at $T=150$\,K, $H=0$\,T. The measured 2D pattern is shown on the top, the calculated is shown on the bottom (a), chi squares normalized per number of points is 5.87. Diffraction profile estimated near the equatorial plane (b). $\gamma$ is azimuthal angle and $\nu$ is elevation angle in the laboratory coodinate system $(xyz)$, where $\vec{x}||\vec{k}_{i}$, $\vec{z}||\vec{H}$\label{fig:(Color-online)-Flipping}.}
\end{figure}

In the absence of a magnetic field the flipping sum diffraction pattern corrected for background is presented in figure\,\ref{fig:(Color-online)-Flipping} together with standard 1D diffraction pattern limited to the equatorial plane. One can see that the scattering intensity distribution along the Debye cones is rather homogeneous and the width of the cones increases with the $\phi$ angle.

\begin{figure}
\includegraphics[width=7cm,height=7cm]{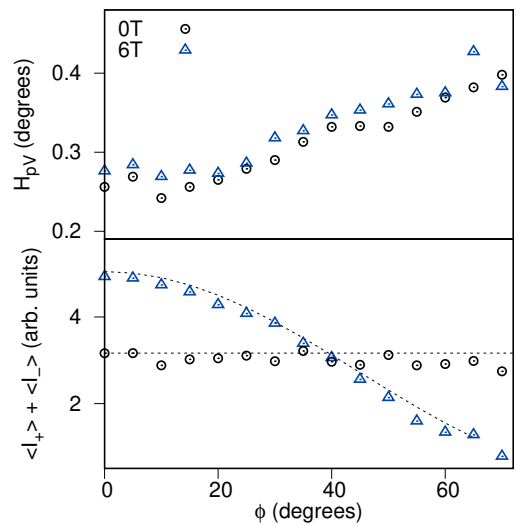}

\caption{(Color online) The distribution of $H_{pV}$ and the sum of the integrated intensities for the reflection $(111)$ measured with the magnetic field 6T (triangles) and without it (circles) at the diffractometer 5C1 at 150\,K over the Debye cone ($\phi=0{^\circ}$ correspond to the scattering in the equatorial plane). The model values after 2D Rietveld refinement are given by the dotted lines.\label{fig:BReflection111}} \end{figure}

Figure\,\ref{fig:BReflection111} shows the $\phi$ dependence of the halfwidth $H_{pV}$ and of the integrated intensity for (111) ($2\theta_{h}=9.42^{\circ}$) reflection. As seen from the figure the width $H_{pV}$ remains approximately constant in the angular range from $0{^\circ}$ to $20^{\circ}$ and strongly increases at higher angles. Therefore the angular range from $0^{\circ}$ to $20^{\circ}$ was used in the refinement. In the meantime we note that the integrated intensity of the (111) reflection remains constant along the whole Debye cone, which is due to the fact that the magnetic moments are randomly oriented.

After refinement by the Rietveld method using equations (\ref{eq:m_perp_sq_aver_smm}, \ref{eq:m_perp_p_aver_smm}) the magnetic moments of iron in tetrahedral and octahedral positions at 150\,K and 0\,T are found to be $-4.23(9)\mu_{B}$, $3.76(6)\mu_{B}$ for 1D data and $-4.09(2)\mu_{B}$, $3.94(2)\mu_{B}$ for 2D ones. Different signs of magnetic moments at two sites correspond to their antiparallel orientation to each other. One can see that the values of magnetic moments are in agreement for both refinements, while a significant decrease of error bars is observed for 2D data. The refined parameters are also in good agreement with literature: $-4.20(3)\mu_{B}$ and $3.97(3)\mu_{B}$\cite{Wright_2002}.

\begin{figure}
\includegraphics[width=8cm,height=8cm]{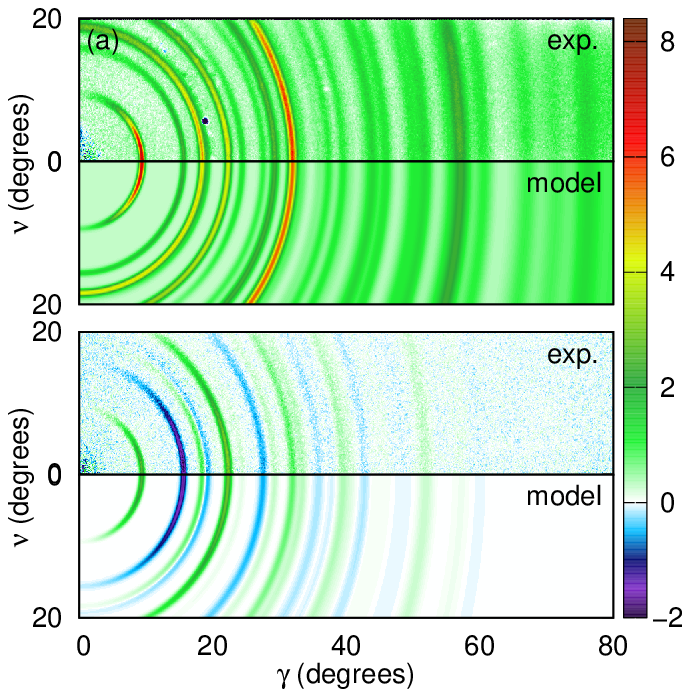}


\includegraphics[width=8cm,height=8cm]{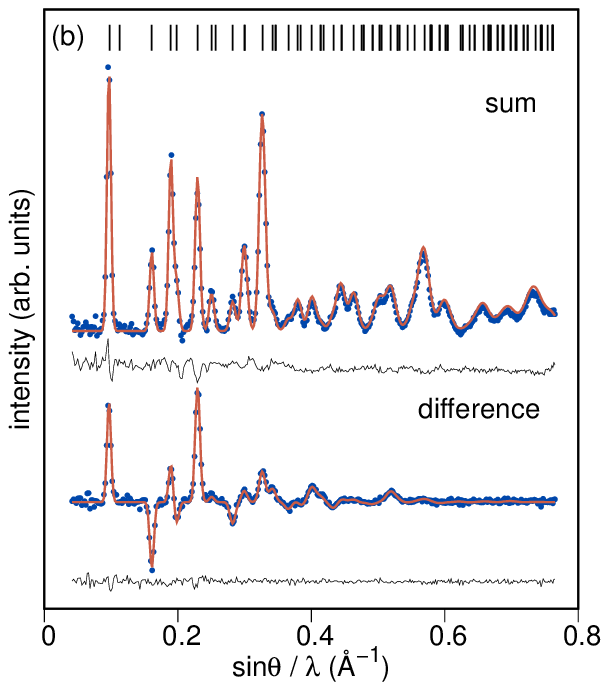}

\caption{(Color online) Flipping sum and difference diffraction patterns collected on Fe\textsubscript{3}O\textsubscript{4} at $T=150$\,K, $6$\,T. The measured 2D pattern is shown on the top, the calculated one is shown on the bottom (a), chi squares normalized per number of points is 3.93. Diffraction profiles estimated near the equatorial plane (b). The position of reflections is marked by ``$|$''. Black lines show the differences between experimental points (blue) and model line (orange).\label{fig:2d-Diffraction-profile-Fe3O4_150K_6T}}
\end{figure}

Flipping sum and difference diffraction patterns measured at 150 K in magnetic field of 6 T are shown in figure\,\ref{fig:2d-Diffraction-profile-Fe3O4_150K_6T}. The presence of magnetic scattering depending on the neutron spin orientation is clearly seen in the flipping difference pattern where reflections with significant magnetic contribution are easily recognizable by strong variation of their intensity along the Debye cone. The angular dependence of the integrated intensity of (111) reflection is shown in figure\,\ref{fig:BReflection111}. It is in a good agreement with the model values calculated by using formula 4 (dotted lines on figure\,\ref{fig:BReflection111}). Note, that strong dependence of magnetic scattering on polar angle $\phi$ rises a problem in the reduction of the two-dimensional diffraction pattern to the one-dimensional one.

Use of polarized neutron diffraction  improve considerably  the precision of the Fe magnetic moments determination. The values of  magnetic moments obtained from the 2D refinement are $-4.03(1)\mu_{B}$  for the ion Fe\textsuperscript{3+} at the tetrahedral site  and $3.95(1)\mu_{B}$ for that at the octahedral site. For the 1D data the moments are found to be $-4.05(7)\mu_{B}$ and $3.89(6)\mu_{B}$, respectively.
Measurements performed below the Verwey transition did not show any evolution of the scattering signal (see Supplemental Material). No new magnetic reflections associated with the ordering of octahedral B irons (Fe\textsuperscript{3+}, Fe\textsuperscript{2+}) were observed, which is in agreement with previous polarized neutron single crystal diffraction measurements\cite{Rakhecha_1978}.

\section{Spin-ice compound Ho\textsubscript{2}Ti\textsubscript{2}O\textsubscript{7}}

Among rare earth pyrochlores titanates Ho\textsubscript{2}Ti\textsubscript{2}O\textsubscript{7} is considered as canonical spin ice compound that shows various exotic magnetic states produced by the presence of geometric frustration\cite{Harris_1997, Gardner_2010}. It shows Ising-like behavior, with the magnetic moments being constrained along the local $\left<111\right>$ axes. In the pyrochlore lattice, distinction between Ising, Heisenberg, or XY models cannot be based, as usual, on the analysis of the macroscopic properties of a single crystalline sample in a magnetic field because of the presence of four different anisotropy axes. The information about the local anisotropy of Ho\textsubscript{2}Ti\textsubscript{2}O\textsubscript{7} has been first obtained by polarized neutron single crystal diffraction based on the so called ``local susceptibility approach''\cite{Cao_2009}. The temperature behavior of the reported local susceptibility tensor has confirmed the Ising character of Ho local anisotropy and was in  perfect agreement with that calculated from the rare earth crystal field parameters. Here we show that the same information about the local susceptibility tensor can be obtained by using 2D Rietveld refinement of polarized neutron powder diffraction patterns.

\begin{figure}
\includegraphics[width=8cm,height=8cm]{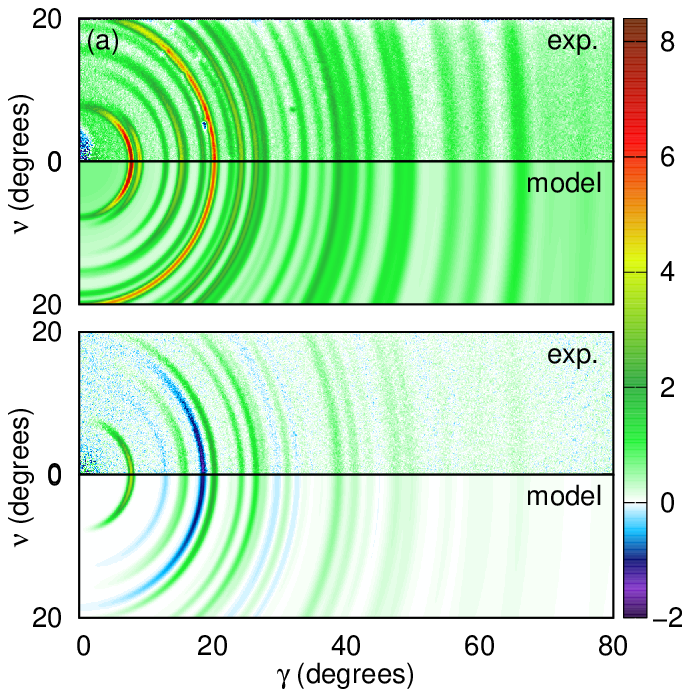}

\includegraphics[width=8cm,height=8cm]{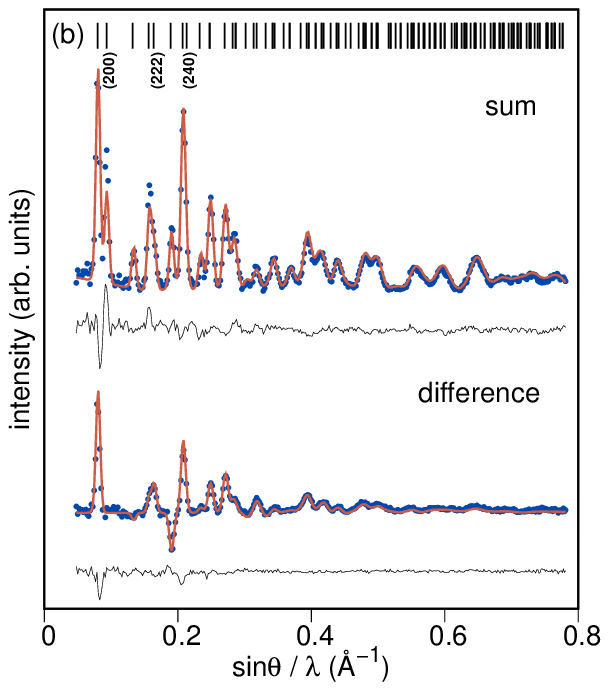}

\caption{(Color online) The measured and calculated flipping sum (top) and difference (bottom) diffraction patterns collected on Ho\textsubscript{2}Ti\textsubscript{2}O\textsubscript{7} at diffractometer 5C1, $T=5$\,K, $H=1$\,T for 2D (a) and 1D (b) diffraction profiles. Chi squares normalized per number of points is 0.57 for 2D diffraction patterns\label{fig:2dRietveld}.}
\end{figure}

We collected a series of powder patterns from the Ho\textsubscript{2}Ti\textsubscript{2}O\textsubscript{7} sintered powder sample in the temperature range 5--50\,K and in field of 1\,T. We found that applying a magnetic field to the sample led to dramatic changes in the diffraction pattern. As an example  the flipping diffraction patterns measured at 5\,K in 1\,T are shown in figure\,\ref{fig:2dRietveld}. As expected a strong variation of intensity along the Debye cone is observed. It can be seen as well that the intensities of reflections allowed by $Fd\overline{3}m$ symmetry (111, 220, 113, etc) are strongly polarization dependent (see the flipping difference pattern in figure\,\ref{fig:2dRietveld}(b)). We also found that the new reflections 200, 222, 240 appear which are forbidden by $Fd\overline{3}m$ symmetry. As seen from the difference plot the intensities of these reflections do not depend on neutron polarization but they are of purely magnetic origin. It has been shown that these reflections arise from the off-diagonal coefficient in the local susceptibility $\chi_{12}$ which becomes significant at low temperatures \cite{Gukasov_2010}. Note that  the  flipping difference pattern, proportional to\,$N\cdot M_{z, \perp}$, contains both positive and negative values, as its sign depends on the phase of the magnetic and nuclear structure factors.

\begin{figure}
\includegraphics[width=7cm,height=7cm]{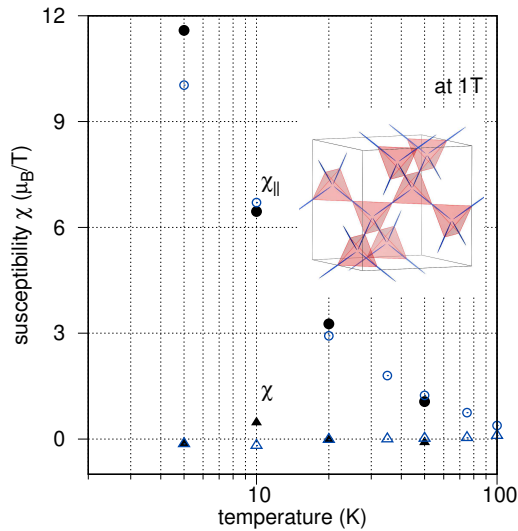}

\caption{(Color online) Temperature dependence of the susceptibility components $\chi_{\parallel}$ (circles) and $\chi_{\perp}$ (triangles) for single (open symbols, taken from\cite{Cao_2009}) and powder (full symbols) Ho\textsubscript{2}Ti\textsubscript{2}O\textsubscript{7} at 1\,T. The insert shows Ho magnetization ellipsoids at 50\,K. \label{fig:(color-online).-Susceptibility}}
\end{figure}

For the space group $Fd\overline{3}m$, the symmetry constraints imply that local susceptibility tensor has only two independent matrix elements $\chi_{11}$and $\chi_{12}$ and the principal axes of Ho magnetization ellipsoids are oriented along the four local $\left<111\right>$ axes. Their lengths given by $\chi_{\parallel}=\chi_{11}+2\chi_{12}$ and $\chi_{\perp}=\chi_{11}-\chi_{12}$ were determined at each temperature. Thermal evolution of $\chi_{\parallel}$ and $\chi_{\perp}$ obtained by 2D Rietveld refinement on polycrystalline sample is shown in the figure\,\ref{fig:(color-online).-Susceptibility} by closed symbols. Open symbols in the figure show the results of previous study performed using polarized neutron diffraction on single crystal\cite{Cao_2009}. One can see that the results of Rietveld refinement are in a good agreement with the single crystal ones and offer the same precision of the susceptibility parameters.

\section{Single-molecule magnet: Co(II) complex}

The 2D Rietveld method is known to be a powerful tool allowing to study powder samples having preferred crystallite orientation\cite{Ferrari1994}. Application of magnetic field to anisotropic powder samples can induce the preferred crystallite orientation, as the net moment of the crystallites tends to align in the field direction. Since the resultant preferred orientation can be determined from the 2D patterns one cans use these ``magnetically textured'' samples in PNPD. We found that such an approach in combination with 2D Rietveld method have a number of advantages and we applied it to the studies of local susceptibility in the cobalt(II) complex with molecular formula Co(L\textsubscript{1})\textsubscript{2}Cl\textsubscript{2}, where L\textsubscript{1} is tetramethylthiourea [(CH\textsubscript{3})\textsubscript{2}N]\textsubscript{2}CS\cite{Vaidya_2018}. The compound is a single-molecule magnet that shows superparamagnetic behavior below a certain blocking temperature and exhibits magnetic hysteresis of purely molecular origin.

\begin{figure}
\includegraphics[width=8cm,height=8cm]{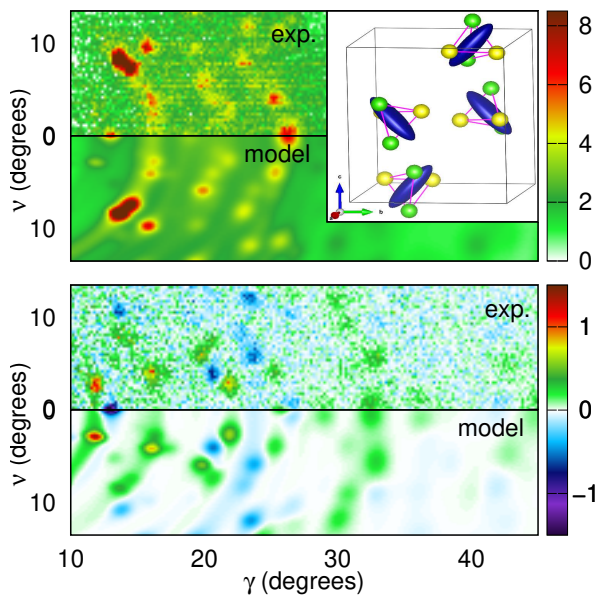}

\caption{(Color online) The measured and calculated flipping sum (top) and difference (bottom) diffraction patterns collected on Co(L\textsubscript{1})\textsubscript{2}Cl\textsubscript{2} at diffractometer 6T2, $T=2$\,K, $H=5$\,T. Chi squares normalized per number of points is 6.71. The insert shows Co magnetization ellipsoids surrounded by S (yellow) and Cl (green).\label{fig:2dRietveld_CoCl2}}
\end{figure}

The powder was filled in a vanadium container of 6\,mm diameter without compressing it. The sample was cooled to 2\,K and the diffraction patterns were measured as a function of magnetic field. The Debye rings in zero field were found homogeneous indicating the absence of preferred crystallite orientation. In magnetic fields above 1\,T the crystallite reorientation started to appear and at 5\,T the Debye rings were transformed in a series of well separated diffraction spots (figure\,\ref{fig:2dRietveld_CoCl2}). We note that the subsequent reduction of the magnetic field to 0\,T did not change the crystallite orientation back to a random one. The diffraction patterns measured at 2\,K and 5\,T in the ``magnetically textured'' sample were used to determine the local susceptibility of the cobalt ion.

To take into account the preferred orientation we used a modified March model\cite{Dollase_1986}, which was developed to describe the mechanism of grain rotation that produces preferred orientation:

\begin{equation}
P_{h}=t+(1-t)\cdot\left[r^{2}\cos^{2}\alpha_{h}+\frac{\sin^{2}\alpha_{h}}{r}\right]^{-3/2},\label{eq:march_function}
\end{equation}

\noindent
where $t$ is the fraction of randomly oriented crystallites, $\alpha_{h}$ is the angle between the transfer momentum and the preferred orientation axis, and $r$ describes the anisotropic shape of crystallites. In the Debye-Scherrer geometry $r$ is more than one for platy crystallites and it is less than one for acicular crystallites. Although in our case the origin of preferred orientation is due to the application of magnetic field, using the March distribution allows estimation of an intuitively simple equivalent specimen compaction.

The studied single-molecule magnet has the monoclinic space group $P2_{1}/n$ with $a=9.88$\,\AA, $b=12.69$\,\AA, $c=14.13$\,\AA, $\beta=92.99^{\circ}$. It is composed of 43 atoms in the asymmetric unit\cite{Vaidya_2018}, including 24 hydrogen atoms. The flipping patterns measured at 2\,K and 5\,T were used to refine the crystalline texture parameters and the susceptibility tensor of cobalt. As seen from figure\,\ref{fig:2dRietveld_CoCl2} a very good agreement between patterns calculated after the refinement and the experimental ones is observed. Both the positions and the widths of the diffraction spots on the Debye cones are well reproduced in the model patterns. The refined March texture parameters $t=0.938(1)$ and $r=0.119(1)$ show that the magnetically induced preferred orientation is rather low at 5 T and only a small part of crystallites is aligned with their easy axes parallel to the field. Hence, stronger fields are needed to overcome the steric hindrence in the powder packing. It is clear, however, that the presence of preferred orientation gives a big advantage in the 2D Rietveld refinement when using area detectors. As seen from the figure different reflections with similar Bragg angles $2\theta$ appear at different $\varphi$ angles. In result no overlapping of these reflections occurs, which allows to use diffractometers with low resolution, like the single crystal diffractometer 6T2, for powder diffraction. We note as well that the conventional approach consisting in the projection of 2D data on the 1D one for Rietveld refinement would result in dramatic decrease of the resolution due to the reflection overlapping. 

Finally, the refined magnetization tensor corresponding to an external field of 5T for cobalt atom in an asymmetric unit was found to be equal to

\begin{equation*}
\left(\begin{array}{ccc}
1.9(3) & 0.0(3) & 0.1(1)\\
0.0(3) & 2.3(3) & -1.4(2)\\
0.1(1) & -1.4(2) & 2.7(3)
\end{array}\right)\mu_{B}.\label{eq:chi_co}
\end{equation*}

\noindent
The corresponding magnetization ellipsoid is presented in the insert of figure\,\ref{fig:2dRietveld_CoCl2}. The averaged magnetization estimated as $2.3(2)$\,$\mu_{B}$ is close to magnetization per cobalt atom $2$\,$\mu_{B}$ taken from the magnetization measurements on polycrystalline sample at 5\,T\cite{Vaidya_2018}. A detailed analysis of the crystallite alignment, the evolution of the magnetization ellipsoids with temperature and field as well as the theoretical interpretation of the ellipsoid orientation are still in progress and will be published later\footnote{J.Overgaard et.al. to be published}.

\section{Conclusion}

Our results suggest that the combination of area detector, 2D Rietveld analysis and the technique of magnetically induced preferred crystallite orientation opens new route to the studies of  local magnetic susceptibility  in polycrystalline materials by  polarized neutron diffraction. The results of 2D Rietveld analysis of diffraction patterns from soft (Fe\textsubscript{3}O\textsubscript{4}) and high (Ho\textsubscript{2}Ti\textsubscript{2}O\textsubscript{7}) magnetic compounds are in  perfect agreement with the single crystal ones reported earlier. We  demonstrate that using  ``magnetically textured'' powder and the 2D Rietveld refinement allows to obtain the precision in the determination of the susceptibility parameters close to that obtained in the single crystal diffraction experiments. By applying this procedure to a single-molecule magnet in polycrystalline form  we obtained the local susceptibility tensor for cobalt atom, which can be now confronted to the theory.  More generally we suggest that the magnetic structure determination  by applying  2D Rietveld method to the ``magnetically textured''   samples has significant perspectives, as it does not require high instrumental resolution due to the fact that different types of reflections with similar Bragg angles are  spread over the Debye cones.  

\section*{Acknowledgments}

We are grateful to J.\,Overgaard for stimulating discussions  and also providing us a single-molecule magnet with Co(II) complex. IK thanks CNRS for his postdoctoral position.

\bibliography{bibliography}

\end{document}